\documentclass[%
reprint,
superscriptaddress,
amsmath,amssymb,
aps,
]{revtex4-2}

\usepackage{graphicx}
\usepackage{dcolumn}
\usepackage{bm}
\usepackage{hyperref}
\usepackage{color} 
\usepackage{orcidlink}

\usepackage{physics}

\newcommand{\atom}[2]{\ensuremath{{}^{#1}\mathrm{#2}}}
\newcommand{\atome}[2]{\ensuremath{{}^{#1}\mathrm{#2}^*}}
\newcommand{\us}{\ensuremath{\mathrm{\mu s}}}
\newcommand{\itsec}[1]{{\textit{#1} }---}

\begin{document}


\title{Neutron multiplicity measurement in muon capture on oxygen nuclei \\ in the Gd-loaded Super-Kamiokande detector}

\newcommand{\AFFicrr}{\affiliation{Kamioka Observatory, Institute for Cosmic Ray Research, University of Tokyo, Kamioka, Gifu 506-1205, Japan}}
\newcommand{\AFFkashiwa}{\affiliation{Research Center for Cosmic Neutrinos, Institute for Cosmic Ray Research, University of Tokyo, Kashiwa, Chiba 277-8582, Japan}}
\newcommand{\AFFicrronly}{\affiliation{Institute for Cosmic Ray Research, University of Tokyo, Kashiwa, Chiba 277-8582, Japan}}
\newcommand{\AFFipmu}{\affiliation{Kavli Institute for the Physics and
Mathematics of the Universe (WPI), The University of Tokyo Institutes for Advanced Study,
University of Tokyo, Kashiwa, Chiba 277-8583, Japan }}
\newcommand{\AFFmad}{\affiliation{Department of Theoretical Physics, University Autonoma Madrid, 28049 Madrid, Spain}}
\newcommand{\AFFubc}{\affiliation{Department of Physics and Astronomy, University of British Columbia, Vancouver, BC, V6T1Z4, Canada}}
\newcommand{\AFFbu}{\affiliation{Department of Physics, Boston University, Boston, MA 02215, USA}}
\newcommand{\AFFuci}{\affiliation{Department of Physics and Astronomy, University of California, Irvine, Irvine, CA 92697-4575, USA }}
\newcommand{\AFFcsu}{\affiliation{Department of Physics, California State University, Dominguez Hills, Carson, CA 90747, USA}}
\newcommand{\AFFcnm}{\affiliation{Institute for Universe and Elementary Particles, Chonnam National University, Gwangju 61186, Korea}}
\newcommand{\AFFduke}{\affiliation{Department of Physics, Duke University, Durham NC 27708, USA}}
\newcommand{\AFFgifu}{\affiliation{Department of Physics, Gifu University, Gifu, Gifu 501-1193, Japan}}
\newcommand{\AFFgist}{\affiliation{GIST College, Gwangju Institute of Science and Technology, Gwangju 500-712, Korea}}
\newcommand{\AFFuh}{\affiliation{Department of Physics and Astronomy, University of Hawaii, Honolulu, HI 96822, USA}}
\newcommand{\AFFicl}{\affiliation{Department of Physics, Imperial College London , London, SW7 2AZ, United Kingdom }}
\newcommand{\AFFkek}{\affiliation{High Energy Accelerator Research Organization (KEK), Tsukuba, Ibaraki 305-0801, Japan }}
\newcommand{\AFFkobe}{\affiliation{Department of Physics, Kobe University, Kobe, Hyogo 657-8501, Japan}}
\newcommand{\AFFkyoto}{\affiliation{Department of Physics, Kyoto University, Kyoto, Kyoto 606-8502, Japan}}
\newcommand{\AFFliv}{\affiliation{Department of Physics, University of Liverpool, Liverpool, L69 7ZE, United Kingdom}}
\newcommand{\AFFmiyagi}{\affiliation{Department of Physics, Miyagi University of Education, Sendai, Miyagi 980-0845, Japan}}
\newcommand{\AFFnagoya}{\affiliation{Institute for Space-Earth Environmental Research, Nagoya University, Nagoya, Aichi 464-8602, Japan}}
\newcommand{\AFFkmi}{\affiliation{Kobayashi-Maskawa Institute for the Origin of Particles and the Universe, Nagoya University, Nagoya, Aichi 464-8602, Japan}}
\newcommand{\AFFpol}{\affiliation{National Centre For Nuclear Research, 02-093 Warsaw, Poland}}
\newcommand{\AFFsuny}{\affiliation{Department of Physics and Astronomy, State University of New York at Stony Brook, NY 11794-3800, USA}}
\newcommand{\AFFokayama}{\affiliation{Department of Physics, Okayama University, Okayama, Okayama 700-8530, Japan }}
\newcommand{\AFFosaka}{\affiliation{Department of Physics, Osaka University, Toyonaka, Osaka 560-0043, Japan}}
\newcommand{\AFFox}{\affiliation{Department of Physics, Oxford University, Oxford, OX1 3PU, United Kingdom}}
\newcommand{\AFFqmul}{\affiliation{School of Physics and Astronomy, Queen Mary University of London, London, E1 4NS, United Kingdom}}
\newcommand{\AFFregina}{\affiliation{Department of Physics, University of Regina, 3737 Wascana Parkway, Regina, SK, S4SOA2, Canada}}
\newcommand{\AFFseoul}{\affiliation{Department of Physics and Astronomy, Seoul National University, Seoul 151-742, Korea}}
\newcommand{\AFFsheff}{\affiliation{School of Mathematical and Physical Sciences, University of Sheffield, S3 7RH, Sheffield, United Kingdom}}
\newcommand{\AFFshizuokasc}{\affiliation{Department of Informatics in
Social Welfare, Shizuoka University of Welfare, Yaizu, Shizuoka, 425-8611, Japan}}
\newcommand{\AFFstfc}{\affiliation{STFC, Rutherford Appleton Laboratory, Harwell Oxford, and Daresbury Laboratory, Warrington, OX11 0QX, United Kingdom}}
\newcommand{\AFFskk}{\affiliation{Department of Physics, Sungkyunkwan University, Suwon 440-746, Korea}}
\newcommand{\AFFtodai}{\affiliation{Department of Physics, University of Tokyo, Bunkyo, Tokyo 113-0033, Japan }}
\newcommand{\AFFtit}{\affiliation{Department of Physics, Institute of Science Tokyo, Meguro, Tokyo 152-8551, Japan }}
\newcommand{\AFFtus}{\affiliation{Department of Physics, Faculty of Science and Technology, Tokyo University of Science, Noda, Chiba 278-8510, Japan }}
\newcommand{\AFFtoronto}{\affiliation{Department of Physics, University of Toronto, ON, M5S 1A7, Canada }}
\newcommand{\AFFtriumf}{\affiliation{TRIUMF, 4004 Wesbrook Mall, Vancouver, BC, V6T2A3, Canada }}
\newcommand{\AFFtokai}{\affiliation{Department of Physics, Tokai University, Hiratsuka, Kanagawa 259-1292, Japan}}
\newcommand{\AFFtsinghua}{\affiliation{Department of Engineering Physics, Tsinghua University, Beijing, 100084, China}}
\newcommand{\AFFynu}{\affiliation{Department of Physics, Yokohama National University, Yokohama, Kanagawa, 240-8501, Japan}}
\newcommand{\AFFllr}{\affiliation{Ecole Polytechnique, IN2P3-CNRS, Laboratoire Leprince-Ringuet, F-91120 Palaiseau, France }}
\newcommand{\AFFbari}{\affiliation{ Dipartimento Interuniversitario di Fisica, INFN Sezione di Bari and Universit\`a e Politecnico di Bari, I-70125, Bari, Italy}}
\newcommand{\AFFnapoli}{\affiliation{Dipartimento di Fisica, INFN Sezione di Napoli and Universit\`a di Napoli, I-80126, Napoli, Italy}}
\newcommand{\AFFroma}{\affiliation{INFN Sezione di Roma and Universit\`a di Roma ``La Sapienza'', I-00185, Roma, Italy}}
\newcommand{\AFFpadova}{\affiliation{Dipartimento di Fisica, INFN Sezione di Padova and Universit\`a di Padova, I-35131, Padova, Italy}}
\newcommand{\AFFkeio}{\affiliation{Department of Physics, Keio University, Yokohama, Kanagawa, 223-8522, Japan}}
\newcommand{\AFFwinnipeg}{\affiliation{Department of Physics, University of Winnipeg, MB R3J 3L8, Canada }}
\newcommand{\AFFkcl}{\affiliation{Department of Physics, King's College London, London, WC2R 2LS, UK }}
\newcommand{\AFFwarwick}{\affiliation{Department of Physics, University of Warwick, Coventry, CV4 7AL, UK }}
\newcommand{\AFFral}{\affiliation{Rutherford Appleton Laboratory, Harwell, Oxford, OX11 0QX, UK }}
\newcommand{\AFFwu}{\affiliation{Faculty of Physics, University of Warsaw, Warsaw, 02-093, Poland }}
\newcommand{\AFFbcit}{\affiliation{Department of Physics, British Columbia Institute of Technology, Burnaby, BC, V5G 3H2, Canada }}
\newcommand{\AFFtohoku}{\affiliation{Department of Physics, Faculty of Science, Tohoku University, Sendai, Miyagi, 980-8578, Japan }}
\newcommand{\AFFicise}{\affiliation{Institute For Interdisciplinary Research in Science and Education, ICISE, Quy Nhon, 55121, Vietnam }}
\newcommand{\AFFilance}{\affiliation{ILANCE, CNRS - University of Tokyo International Research Laboratory, Kashiwa, Chiba 277-8582, Japan}}
\newcommand{\AFFibs}{\affiliation{Center for Underground Physics, Institute for Basic Science (IBS), Daejeon, 34126, Korea}}
\newcommand{\AFFglasgow}{\affiliation{School of Physics and Astronomy, University of Glasgow, Glasgow, Scotland, G12 8QQ, United Kingdom}}
\newcommand{\AFFoecu}{\affiliation{Media Communication Center, Osaka Electro-Communication University, Neyagawa, Osaka, 572-8530, Japan}}
\newcommand{\AFFminn}{\affiliation{School of Physics and Astronomy, University of Minnesota, Minneapolis, MN  55455, USA}}
\newcommand{\AFFsilesia}{\affiliation{August Che\l{}kowski Institute of Physics, University of Silesia in Katowice, 75 Pu\l{}ku Piechoty 1, 41-500 Chorz\'{o}w, Poland}}
\newcommand{\AFFtoyama}{\affiliation{Faculty of Science, University of Toyama, Toyama City, Toyama 930-8555, Japan}}
\newcommand{\AFFbmcc}{\affiliation{Science Department, Borough of Manhattan Community College / City University of New York, New York, New York, 1007, USA.}}

\AFFicrr
\AFFkashiwa
\AFFicrronly
\AFFmad
\AFFbmcc
\AFFbu
\AFFbcit
\AFFuci
\AFFcsu
\AFFcnm
\AFFduke
\AFFllr
\AFFgifu
\AFFgist
\AFFglasgow
\AFFuh
\AFFibs
\AFFicise
\AFFicl
\AFFbari
\AFFnapoli
\AFFpadova
\AFFroma
\AFFilance
\AFFkeio
\AFFkek
\AFFkcl
\AFFkobe
\AFFkyoto
\AFFliv
\AFFminn
\AFFmiyagi
\AFFnagoya
\AFFkmi
\AFFpol
\AFFsuny
\AFFokayama
\AFFoecu
\AFFox
\AFFral
\AFFseoul
\AFFsheff
\AFFshizuokasc
\AFFsilesia
\AFFstfc
\AFFskk
\AFFtohoku
\AFFtokai
\AFFtodai
\AFFipmu
\AFFtit
\AFFtus
\AFFtoronto
\AFFtoyama
\AFFtriumf
\AFFtsinghua
\AFFwu
\AFFwarwick
\AFFwinnipeg
\AFFynu

\author{S.~Miki}
\AFFicrr
\author{K.~Abe}
\AFFicrr
\AFFipmu
\author{S.~Abe}
\AFFicrr
\author{Y.~Asaoka}
\AFFicrr
\AFFipmu
\author{C.~Bronner}
\author{M.~Harada}
\AFFicrr
\author{Y.~Hayato}
\AFFicrr
\AFFipmu
\author{K.~Hiraide}
\AFFicrr
\AFFipmu
\author{K.~Hosokawa}
\AFFicrr
\author{K.~Ieki}
\author{M.~Ikeda}
\AFFicrr
\AFFipmu
% ...


\collaboration{The Super-Kamiokande Collaboration}
\noaffiliation

\date{\today}

\begin{abstract}
In recent neutrino detectors, neutrons produced in neutrino reactions play an important role. Muon capture on oxygen nuclei is one of the processes that produce neutrons in water Cherenkov detectors. We measured neutron multiplicity in the process using cosmic ray muons that stop in the gadolinium-loaded Super-Kamiokande detector. For this measurement, neutron detection efficiency is  obtained with the muon capture events followed by gamma rays to be $50.2^{+2.0}_{-2.1}\%$. By fitting the observed multiplicity considering the detection efficiency, we measure neutron multiplicity in muon capture as $P(0)=24\pm3\%$, $P(1)=70^{+3}_{-2}\%$, $P(2)=6.1\pm0.5\%$, $P(3)=0.38\pm0.09\%$. This is the first measurement of the multiplicity of neutrons associated with muon capture on oxygen without neutron energy threshold.
\end{abstract}


\maketitle

\itsec{Introduction}
In recent neutrino detectors, neutrons play an important role in reducing background events in searches for diffuse supernova background~\cite{Harada2023} or nucleon decay~\cite{Takenaka2020a} and in distinguishing neutrinos and anti-neutrinos in neutrino oscillation analyses~\cite{Wester2024}.
To improve neutrino analyses, it is crucial to understand the mechanisms of the neutron production and their multiplicities.

The muon capture reaction on an oxygen nucleus 
\begin{equation*}
    \mu^- + \atom{16}{O} \rightarrow \nu_\mu + A + n + n + \cdots
\end{equation*}
is one of the processes that produce neutrons after neutrino reactions in water Cherenkov detectors.
Here, $\nu_\mu$, $n$, and $A$ denote muon neutrino, neutron, and the residual nucleus, respectively.
After negative muons lose their energy in water, they are trapped by a nearby oxygen nucleus and $18.4\pm0.1$\% of them are captured on the nucleus~\cite{Suzuki1987}. 
Captures on hydrogen are negligible because muonic hydrogen $\mu^- p$ is a neutral system and easily penetrates to a nearby nucleus, and the muon is transferred to another nucleus with a larger atomic number with stronger binding energy~\cite{Measday2001}. 

In the perspective of nuclear physics, neutron multiplicity in the muon capture reaction is governed by the excitation function of nucleons. 
Measuring neutron multiplicity helps us to estimate the excitation function and to study the nucleon momentum distribution in oxygen nuclei.

Muon capture reactions have been studied since as early as 1950's~\cite{Kaplan1958, Winsberg1954}.
Recently, for heavier nuclei such as molybdenum, neutron emission in muon capture reactions has been measured  for double beta decay experiments~\cite{Hashim2018, Hashim2020, Othman2021a}. The observed multiplicity can be described by the pre-equilibrium -- equilibrium model, which treats a large number of nucleons in the statistical mechanical way, but this prescription is not applicable to light nuclei like oxygen.

Particle emissions in muon capture reactions on oxygen were actively studied in the 1970-1980's as reviewed in~\cite{Measday2001}.
When negative muons are captured on \atom{16}{O}, $10.9\pm1.2\%$ result in \atom{16}{N}~\cite{Kane1973}, $20\pm6\%$ in \atom{15}{N} ground state with one neutron emission, $39\pm7\%$ in \atom{15}{N} 6.32\,MeV excited state with one neutron emission, $0.8\pm0.4\%$ in \atom{14}{N} 3.9\,MeV excited state with two neutron emissions~\cite{VanDerSchaaf1983}.
Here, the \atom{15}{N} ground-state yield was deduced by adding the yields with a 5\,MeV and 6.5\,MeV neutron~\cite{VanDerSchaaf1983}.
However, the branching ratios have been measured by coincident measurement of neutrons and de-excitation gamma rays, and no experiment has measured the neutron multiplicity directly.
Yields of the residual nucleus are recently measured in this reaction~\cite{maekawa2025}.

In this article, we report the result of neutron multiplicity measurement in the muon capture reaction on \atom{\mathrm{nat}}{O} without any requirement of coincident de-excitation gamma rays. 
We use the gadolinium-loaded Super-Kamiokande (SK) detector as a neutron detector~\cite{Abe2022}. 
It has $4\pi$ spatial coverage and is sensitive to a wide range of neutron energy, thermal to GeV, without a lower energy threshold, so it enables unbiased multiplicity measurement independent of neutron energy or direction.
We use the cosmic ray muons that stopped in the detector as the source.
Such events are frequently recorded during the normal data taking of SK as a neutrino detector.

\itsec{Detector and data}
The SK detector is a 50-kton water Cherenkov detector located 1000~m underground in the Kamioka mine in Japan~\cite{Fukuda2003, Abe2014}.
A cylindrical tank of water is optically separated into an inner detector (ID) and an outer detector (OD) surrounding the ID. 
The total volume of the ID is 32~kton and it is observed by over 11,000 inward photomultiplier tubes (PMTs) on the wall, while the two-meter-thick OD works as an active veto for cosmic ray muons and as a shield for incoming neutrons and gamma rays. 
Relativistic charged particles emit cones of Cherenkov photons in the detector.
Charge and timing information of photons from the Cherenkov radiation observed by ID PMTs enable us to reconstruct event vertex, direction, and momentum, and to classify the event as e-like for electrons and gamma rays or $\mu$-like for muons and charged pions.

In 2020, gadolinium (Gd) sulfate was dissolved into the detector's water to achieve 0.011\% Gd concentration in order to improve neutron detection efficiency~\cite{Abe2022}.
With this concentration, half of thermal neutrons in the detector are captured on Gd with a subsequent 8~MeV gamma ray cascade, while the other half are captured on free protons with a 2.2~MeV gamma ray.
By detecting gamma rays following primary events, neutrons can be identified with an efficiency of $50.2^{+2.0}_{-2.1}\%$.
This article also reports how this efficiency is obtained.

The analysis begins with selecting and reconstructing cosmic ray muon events which stopped in the ID (stopping muons).
Events are selected as stopping muons when there is only one hit cluster in the OD PMTs corresponding to the muon entrance point. 
The entrance point, direction, and momentum of the muon are reconstructed from the observed charge distribution and the hit timing in the ID PMTs. 
Assuming energy loss of muons in water~\cite{PDG2020}, the stopping point of the muon is reconstructed. 
From the simulated stopping muon events, the resolution of the stopping point is estimated to be 102~cm.
To ensure that neutrons emitted at the stopping point are captured by Gd or protons in the inner region of the ID, we required that the reconstructed stopping point be more than 300~cm away from the ID wall. 
In the 562.4 live-days of data from 2020 to 2022 that we use in the analysis, 1,986,465 events are selected as stopping muons, corresponding to 3,532~events/day.

\itsec{Neutron efficiency measurement}
For the estimation of the neutron tagging efficiency, we use stopping muon events followed by high-energy gamma rays.
Among the decay branches of the \atome{16}{N} formed in muon capture reactions, those accompanied by de-excitation gamma rays above several MeV are \atom{15}{N} with one neutron emission ($67\pm 8\%$~\cite{VanDerSchaaf1983}) and \atom{14}{N} with two neutron emissions (estimated roughly 0.8\%~\cite{Measday2001}). 
Here, the fractions in the parentheses show the branching ratios among total muon captures on \atom{16}{O}.
Other states in \atom{16}{N} and \atom{14}{N} have lower excited energies.
Especially, the lowest excited state of \atom{16}{N} above 1~MeV is at 7.5~MeV, but this is high enough to emit a neutron, and the 7.5~MeV gamma ray is not observed from the \atome{16}{N} state.
Thus, we define a "control sample" as a subsample of the muon capture events with a high-energy gamma ray signature, predominantly corresponding to single-neutron emission. Though there is a small contamination from other channels, it provides a reasonably pure single-neutron reference.
We use events with gamma rays with more than 30 PMT hits in ID as a reference sample. 
This threshold corresponds to 5~MeV of visible energy.

\begin{figure}
\includegraphics[scale=0.4]{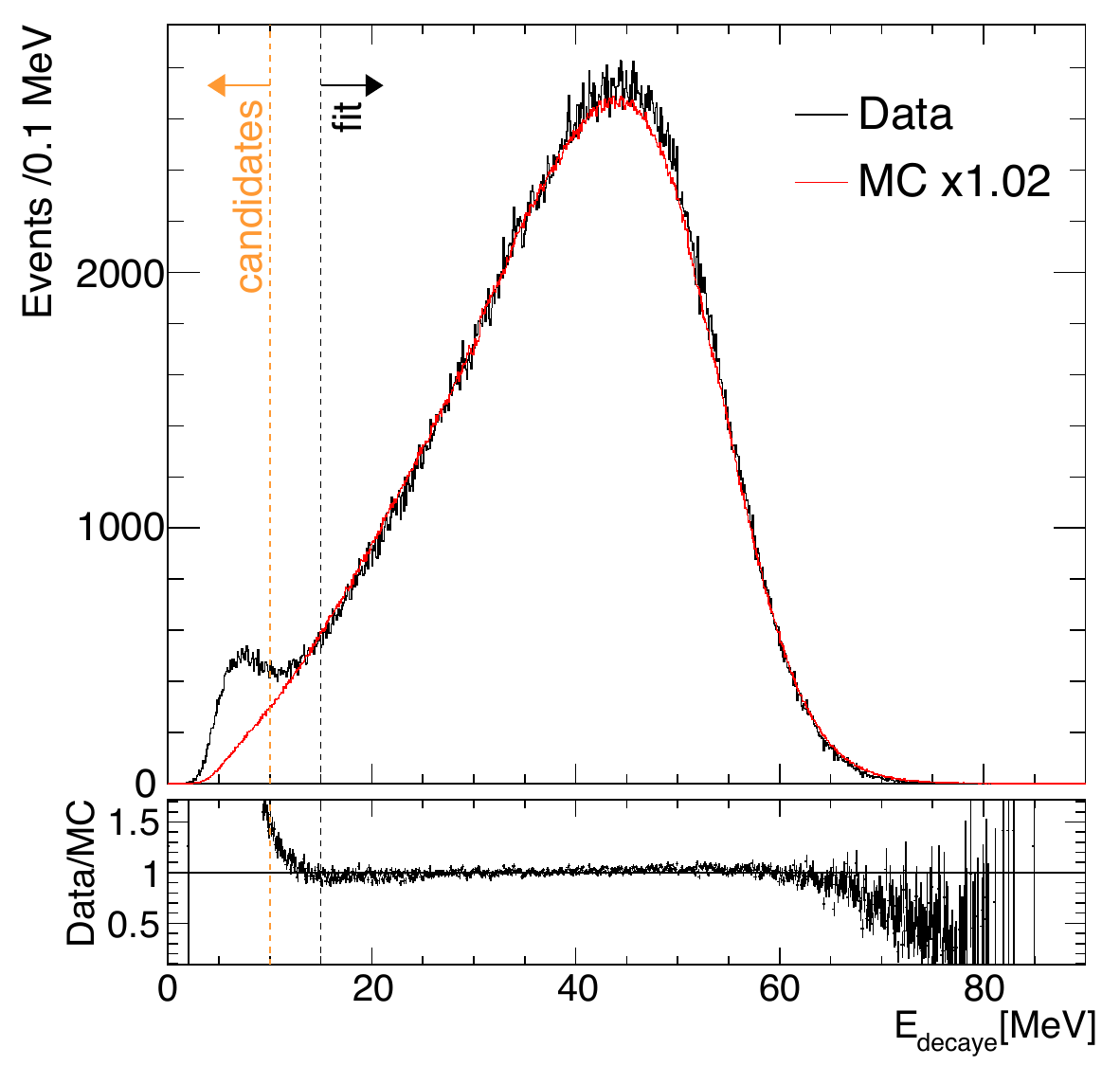}
\caption{\label{fig:decaye_spectrum_with_MC}
The observed and simulated energy spectrum of decay electrons and de-excitation gamma rays following stopping muons. The time difference between the decay electron or the gamma ray and the preceding stopping muon is required to be within [1.1, 5]~\us . The simulation includes only decay electrons and is normalized with the number of stopping muons. The energy in MC is scaled by $+2.0\%$ to match the spectrum in data. The fit includes events with energies greater than 15 MeV.}
\end{figure}

De-excitation gamma rays after muon capture reactions are observed in the same way as decay electrons, but their energies are lower than those of decay electrons.
Figure~\ref{fig:decaye_spectrum_with_MC} shows the reconstructed energy spectrum of decay electrons (for MC) and of decay electrons and de-excitation gamma rays (for data). 
The signals are required to be detected within [1.1, 5]~\us\ after the stopping muon.
At shorter time scales, the spectrum is distorted by the dead time of the data acquisition system, which is 900~ns after a PMT hit~\cite{Nishino2009}.
At longer time scales, false-tagged signals which are expected to be flat in time account for a larger proportion of the signal.
The energy in MC is scaled by $+2.0\%$ to match the observed energy spectrum. 
This $2.0\%$ is consistent with the systematic energy uncertainty evaluated for other analyses~\cite{Wester2024}. 

Using the energy spectrum, the events below 10~MeV are selected as gamma ray candidates. 
There are $N_{\mathrm{can}}=26,454$ candidate events selected in data, and $N_{\mathrm{decay}}=9,425$ decay electrons are expected to contaminate the candidates from the MC spectrum.

Other possible sources of contamination originating from stopping muons are radioactive isotopes and neutron capture signals, both of which are longer lived than 100~\us\ after the parent muons.
Background signals independent of stopping muons, such as radioactive background~\cite{Nakano2020} and spallation products~\cite{Zhang2016}, are expected to occur accidentally.
Thus, background contamination other than decay electrons is estimated as a flat component in the decay time distribution of the candidate events. 
Among the candidate events, the number of events originating from $\mu^+$ is calculated as $N_+ = N_{\mathrm{decay}} \frac{r}{1+r-p_\mathrm{cap}}$ where $r=1.32\pm0.02$ is the charge ratio of cosmic ray muons~\cite{Kitagawa2024} and $p_\mathrm{cap}=0.184\pm0.001$ is the probability that a $\mu^-$ is captured on a nucleus~\cite{Suzuki1987}. 
The other events are assumed to be originating from $\mu^-$: $N_- = N_{\mathrm{can}}-N_+$. 
Then the decay time distribution of the candidate events is fitted with 
\begin{equation}
     p_0 \qty[\frac{N_+}{\tau^+} \exp (-\frac{t
     }{\tau^+}) + \frac{N_-}{\tau^-} \exp (-\frac{t}{\tau^-})] + p_1
\end{equation}
where $\tau^\pm$ are the lifetimes of $\mu^\pm$, fixed at $\tau^+=2.197~\us$ and $\tau^-=1.795~\us$ in oxygen~\cite{Suzuki1987}, and $p_0$ and $p_1$ are free parameters. 
The number of contaminating events is estimated to be $N_{\mathrm{const}}=621$ from the best-fit value of $p_1$.

The candidate event sample contains $N_{\gamma}=N_{\mathrm{can}}-N_{\mathrm{decay}}-N_{\mathrm{const}}=16,408$ de-excitation gamma rays from \atome{15}{N}, and $N_{\gamma}$ is expected to be equal to the number of neutrons emitted in the candidate event sample because muon capture events followed by high energy de-excitation gamma rays form a single-neutron control sample.

\begin{figure}
\includegraphics[scale=0.4]{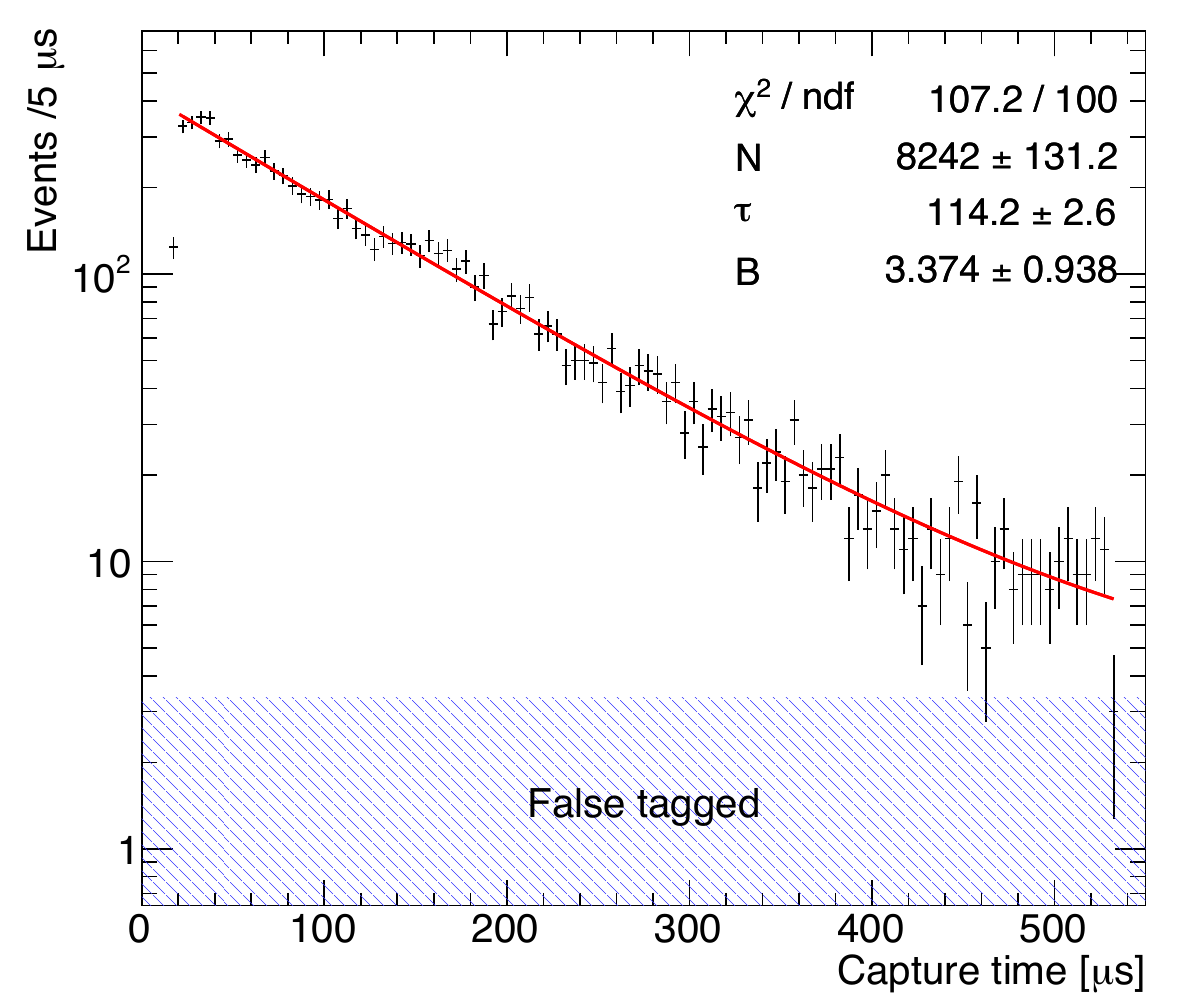}
\caption{\label{fig:ntag_capture_time}
The detection time of neutrons with respect to the preceding stopping muons in those events with de-excitation gamma ray candidates. The search window is [18, 535]~\us\ after the stopping muons to avoid after-pulse of PMTs, which happens 12--18~\us\ after a large hit~\cite{Abe2022a}. The red line shows the fitting result by Eq.~\eqref{eq:captue_time_fit}. The hatched box represents the $B$ component, corresponding to the false tagged signals.}
\end{figure}

In the candidate events, 8,568 neutron signals are tagged by the method by~\cite{Han2025,Han2023}.
The method searches for PMT hit clusters following muon events, and applies a neural network trained with simulated neutron signals to identify neutrons.
Fig.~\ref{fig:ntag_capture_time} shows the distribution of the neutron detection time with respect to stopping muons. 
The distribution is fitted with 
\begin{equation}
    \frac{N\times5~\us}{\tau(e^{-18~\us /\tau}-e^{-535~\us /\tau})}e^{-t/\tau}+B
    \label{eq:captue_time_fit}
\end{equation}
with $\tau$, $N$, and $B$ as fitting parameters, where $\tau$ is the time constant of the neutron capture reaction.
Here, $5~\us$ corresponds to the bin width of Fig.~\ref{fig:ntag_capture_time}, and $18~\us$ and $535~\us$ are the start and end times of the neutron search window after the stopping muon.
By integrating the function over the search time window, $N$ gives the number of the detected neutron signals, and $B$ represents time-independent backgrounds which are made up of the false tagged signals.
The fitted value of $\tau=114.2\pm2.6~\us$ is consistent with measurements with an americium/beryllium neutron source~\cite{Abe2022}.

The number of the detected neutrons is obtained as $N=8,242$, with which the neutron tagging efficiency is calculated by $N/N_\gamma$.

The effect of energy scaling by 2.0\% of the decay electron spectrum in MC in Fig.~\ref{fig:decaye_spectrum_with_MC} is treated as systematic uncertainty.
To estimate the number of de-excitation gamma rays, the time-independent contamination events, $N_\mathrm{const}$, are discarded because the tagged events are not de-excitation gamma rays. 
However, in those events, decay electrons are not detected as well, so they can be muon capture events and they can be accompanied by neutrons.
Therefore, we should not ignore those events in calculating $N_\gamma$, and the number of those discarded events, $N_\mathrm{const}$, is considered as a systematic uncertainty of $N_\gamma$.

This analysis assumes that muon capture events with $>5$~MeV gamma rays are accompanied by one neutron.
Previous measurements of neutrons and gamma rays from the muon capture reaction reported that 0.8\% of the reaction is associated with gamma rays with a total energy of 3.9~MeV and two neutrons~\cite{VanDerSchaaf1983}. 
From the similarity in branching ratios in $\atom{16}{O}(\mu^- , \nu nn)$ and $\atom{16}{O}(\gamma , pn)$ reactions, it is estimated that the branching ratios to the 3.9~MeV state and 7.0~MeV state of \atom{14}{N} are roughly equal in both reactions~\cite{Measday2001}. 
Hence 0.8\% of muon capture events are estimated to contain two neutrons and 7.0~MeV gamma rays, though the gamma rays are confused with the numerous \atom{15}{N} gamma rays in the measurements.
This branch can cause bias in the efficiency measurement in this analysis, and the ratio is considered as a systematic uncertainty in the efficiency.
The fraction of this branching ratio, $Br(\atom{14}{N}(7.0~\mathrm{MeV}))/Br(\atom{15}{N}~\mathrm{excited~states})=1.2\%$, is set as a relative uncertainty of the detection efficiency.

\begin{table}[b]
\caption{\label{tab:efficiency_syst}
Summary of the systematic uncertainties of the measured neutron detection efficiency.
The uncertainty sizes are expressed as the absolute uncertainty in the detection efficiency.
}
\begin{ruledtabular}
\begin{tabular}{lc}
\multicolumn{1}{c}{Source}&Uncertainty\\
\colrule
Energy scaling & $+1.9\%$\\
Time-independent contamination to $\gamma$ & $-1.9\%$\\
Capture time fit (including statistics) & $\pm0.8\%$\\
$\atom{16}{O}(\mu^- , \nu nn)\atom{14}{N}$ reaction & $-0.6\%$\\
Muon charge ratio, $r$ & $<0.01\%$\\
Capture probability, $p_\mathrm{cap}$ & $<0.001\%$\\
Lifetime of $\mu^\pm$ & $\pm0.02\%$\\
Distortion by neutron thermalization & $\pm0.02\%$\\
\colrule
Total & $+2.0/-2.1\%$
\end{tabular}
\end{ruledtabular}
\end{table}

Other uncertainties are summarized in Table~\ref{tab:efficiency_syst}.
The neutron tagging efficiency in the muon capture reaction is measured to be $50.2^{+2.0}_{-2.1}\%$ with statistical and systematic uncertainties.
This efficiency is stable throughout the observation period and is uniform in any region in the fiducial volume in the detector.

\itsec{Neutron multiplicity measurement}
As the next step, we measure neutron multiplicity in the muon capture process. 
We use the same data set as we used in the efficiency estimation. 
This time, to collect all muon capture events without any bias from de-excitation gamma rays, we remove stopping muon events with decay electrons whose energy is higher than 15~MeV. 
Here, we assume that no muon capture results in more than 15~MeV of de-excitation gamma rays.
From the MC sample, the selection efficiency of this cut for muon capture events is estimated to be 99.88\%.

The distribution of the number of detected neutrons in the selected events is shown in the second row in Table~\ref{tab:multiplicity}. 
\begin{table}[b]
\caption{\label{tab:multiplicity}
The numbers of neutron signals in a event.
The second row shows the observed distribution in the events after the decay electron cut in a search window of [18, 535]~\us\ after the preceding muons.
The third row shows the fitted distribution assuming Eqs.~\eqref{eq:correction1} and \eqref{eq:correction2}, which is considered to be total generated neutrons.
The fourth row shows the distribution in muon capture events obtained by subtracting contribution of decay events.
}
\begin{ruledtabular}
\begin{tabular}{crrrrrrr}
Number of neutrons& 0 & 1 & 2 & 3 & 4 & 5 & Sum\\
\colrule
Observed & 816,249 & 66,514 & 3,197 & 112 & 5 & 2 & 886,079 \\
Fitted & 766,292 & 109,527 & 9,574 & 592 & 0 & 32 & 886,018 \\
In capture events & 37,821 & 109,527 & 9,574 & 592 & 0 & 32 & 157,547\\
\end{tabular}
\end{ruledtabular}
\end{table}
PMT hits from each neutron capture are distributed over a time of order 100~ns while the capture time constant is around 100~\us, so the detected neutrons can be assumed to be statistically independent.
The multiplicity distribution is distorted by tagging efficiency $\varepsilon$ and false tagged signals whose rate is $p$ in units of $\mathrm{event}^{-1}$.
When we denote the number of events where $i$ neutrons are generated as $n_i$, the number of events with $i$ tagged neutrons is 
\begin{equation}
    t_i = \sum_{j=i}^\infty n_j C(j,i) \varepsilon^i (1-\varepsilon)^{j-i}
    \label{eq:correction1}
\end{equation}
where $C(j,i)$ is the number of $i$-combinations from $j$ elements. By adding false tagged signals, the expected number of events with $i$ observed signals is written as
\begin{equation}
    E_i(n) = t_i (1-p) + \sum_{j=0}^{i-1} t_j p^{i-j}
    \label{eq:correction2}
\end{equation}
as a function of assumed true multiplicity $n$.
The number of detected false neutron signals is obtained from the fitted $B$ in Eq.~\eqref{eq:captue_time_fit} of neutron detection time distribution, and the false tagging rate is obtained as $p=$0.0083(4) per event.

The observed multiplicity $O_i$ is unfolded to obtain the generated multiplicity $n_i$ by minimizing a $\chi^2$ statistic defined as
\begin{equation}
\begin{aligned}
    \chi^2 = 2 \sum_{i=0}^5 \qty(E_i(n) - O_i + O_i \log{\frac{O_i}{E_i(n)}}) \\
           + \qty(\frac{\varepsilon - \varepsilon_0}{\sigma_\varepsilon})^2 \
           + \qty(\frac{p - p_0}{\sigma_p})^2 
\end{aligned}
\end{equation}
where $\varepsilon_0$ and $p_0$ ($\sigma_\varepsilon$ and $\sigma_p$) are the estimated pre-fit values (uncertainties) of $\varepsilon$ and $p$ respectively.
The statistic compares the expected and observed multiplicity assuming Poisson fluctuations with two systematic uncertainties as nuisance parameters.

\begin{figure}
\includegraphics[width=\linewidth]{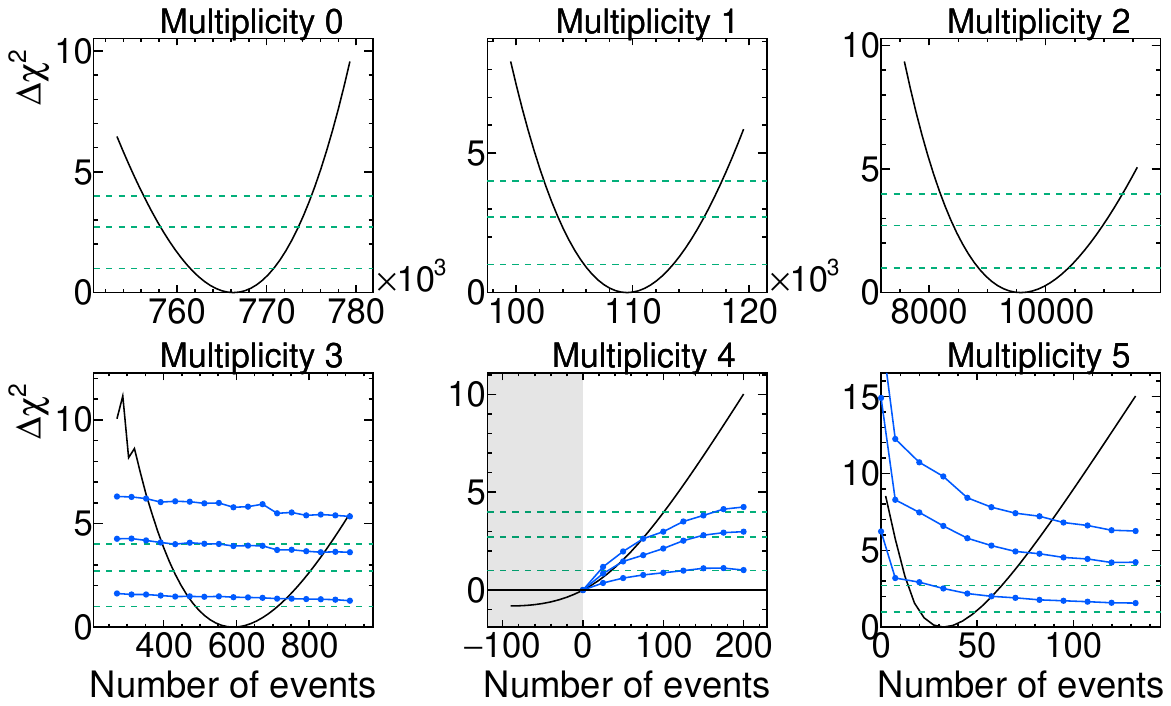}
\caption{\label{fig:fitted_chi2_with_FC}
$\Delta\chi^2$ contours for the number of events with each neutron multiplicity obtained by unfolding the observed distribution. Blue lines with dot markers show the Feldman-Cousins critical values~\cite{Feldman1998} for $1\sigma$, 90\%, and $2\sigma$ confidence level from bottom to top. Green dashed lines represent $\Delta \chi^2 = 1,\ 2.71$, and $4$, roughly corresponding to $1\sigma$, 90\%, and $2\sigma$ confidence levels, from bottom to top.}
\end{figure}

The $\chi^2$ is minimized against $n_i$, $\varepsilon$, and $p$ while all the fitting parameters are constrained to be equal to or greater than zero.
Figure~\ref{fig:fitted_chi2_with_FC} shows the one-dimensional $\Delta\chi^2$ contours of the parameters, obtained as the difference between the global minimum $\chi^2$ and the minimum $\chi^2$ with one parameter of interest fixed.
The best-fit values of $n_i$ are listed in the third row in Table~\ref{tab:multiplicity}.
For $i\ge 3$, the fitted parameter regions are close to the unphysical regions, so uncertainties of the parameters are obtained by the Feldman-Cousins approach~\cite{Feldman1998}.
The critical values are overlaid in Fig.~\ref{fig:fitted_chi2_with_FC}.
The fit was performed up to $n_5$ here, but including $n_6$ does not lead to significant changes in the fit result.

The fitted number of events with 0 neutrons contains stopping muon events where the muon is not captured but decays into an electron. 
The number of muon capture events is calculated as
\begin{equation}
    N_\mathrm{cap} = N_\mu\ \frac{1}{1+r}\ p_\mathrm{cap} = 157,547
\end{equation}
where $N_\mu$ is the number of stopping muon events, $r$ is the muon charge ratio, and $p_\mathrm{cap}$ is the capture probability of $\mu^-$. 
We decrease the number of events with 0 neutrons so that the total number of events agrees with $N_\mathrm{cap}$. 
In this way, we obtain the $n_i$ distribution in muon capture events as in the fourth row in Table~\ref{tab:multiplicity}.

As we described earlier, the muon capture process on hydrogen \atom{1}{H} can be ignored, but we have gadolinium and sulfur atoms in the detector water. 
In chemical compounds, muon Coulomb capture probability is calculated as the number ratio of nuclei multiplied by relative capture probability $R(Z)$, where $Z$ is the atomic number of the nuclei. 
$R(Z)$ was experimentally measured \cite{VonEgidy1982}.
In the detector water with 0.021\% $\mathrm{Gd_2(SO_4)_3}$, 0.007\% of stopping negative muons are Coulomb captured by gadolinium, and 0.002\% are Coulomb captured by sulfur. 
Among the 1,986,465 stopping muons in our sample, 62 and 20 $\mu^-$ are estimated to be captured on gadolinium and sulfur respectively.
Neutron multiplicity in muon captures on these nuclei is uncertain, so conservatively we set the sum of these numbers, 82, for a systematic uncertainty of the numbers in the fourth row in Table~\ref{tab:multiplicity}.
With this uncertainty, the numbers of events of multiplicities four and five cannot be distinguished from those of captures on gadolinium or sulfur.

As a result, the probability $P(i)$ of $i$ neutrons emitted in the muon capture process on oxygen of natural abundance is calculated to be $P(0)=24\pm3\%$, $P(1)=70^{+3}_{-2}\%$, $P(2)=6.1\pm0.5\%$, $P(3)=0.38\pm0.09\%$. 
Systematic uncertainties from the charge ratio $r(\mu^+/\mu^-)=1.32\pm0.02$ of cosmic ray muons \cite{Kitagawa2024}, and the capture probability of $\mu^-$ $p_\mathrm{cap}=18.4\pm0.1\%$ \cite{Suzuki1987} are considered in addition to the fitting uncertainty coming from statistics, neutron detection efficiency, and neutron false tagging rate.
The uncertainties are evaluated from the variance of $P(i)$ resulting from varying $r$ and $p_\mathrm{cap}$.
The uncertainties of $r$ and $p_\mathrm{cap}$ are correlated, but the correlation is estimated to be a factor of 20 smaller than the total uncertainty of $r$, so this correlation is ignored in this article.

$P(2)=6.1\pm0.5\%$ measured in this study is larger than in the previous measurement \cite{VanDerSchaaf1983, Measday2001}, where neutrons with energies greater than 0.9~MeV were measured.
It suggests that more neutrons are emitted at lower energy in muon capture reactions.

\itsec{Summary}
We have measured neutron multiplicity for up to three neutron emissions in the muon capture reaction on oxygen nuclei of natural abundance with stopping muon events in the gadolinium-loaded Super-Kamiokande detector.
For this measurement, neutron detection efficiency has been obtained to be $50.2^{+2.0}_{-2.1}\%$ by using capture events with gamma rays as a one-neutron control sample. 
This is the first study to measure neutron multiplicity in this reaction directly without a neutron energy threshold in the reaction.

Our results will improve neutron simulation in water Cherenkov detectors and therefore contribute to more precise neutrino oscillation measurements and rare event searches. 
In addition, the precise measurement of the neutron multiplicity gives the excitation function in the process and could help to study nucleon momentum distribution in nuclei.

\itsec{Acknowledgments}
We gratefully acknowledge the cooperation of the Kamioka Mining and Smelting Company.
The Super-Kamiokande experiment has been built and operated from funding by the 
Japanese Ministry of Education, Culture, Sports, Science and Technology; the U.S.
Department of Energy; and the U.S. National Science Foundation. Some of us have been 
supported by funds from the National Research Foundation of Korea (NRF-2009-0083526,
NRF-2022R1A5A1030700, NRF-2022R1A3B1078756) funded by the Ministry of Science, 
Information and Communication Technology (ICT); the Institute for 
Basic Science (IBS-R016-Y2); and the Ministry of Education (2018R1D1A1B07049158,
2021R1I1A1A01042256, 2021R1I1A1A01059559, RS-2024-00442775);
the Japan Society for the Promotion of Science; the National
Natural Science Foundation of China under Grants No.12375100; the Spanish Ministry of Science, 
Universities and Innovation (grant PID2021-124050NB-C31); the Natural Sciences and 
Engineering Research Council (NSERC) of Canada; the Scinet and Westgrid consortia of
Compute Canada; 
the National Science Centre (UMO-2018/30/E/ST2/00441 and UMO-2022/46/E/ST2/00336) 
and the Ministry of  Science and Higher Education (2023/WK/04), Poland;
the Science and Technology Facilities Council (STFC) and
Grid for Particle Physics (GridPP), UK; the European Union's 
Horizon 2020 Research and Innovation Programme under the Marie Sklodowska-Curie grant
agreement no.754496; H2020-MSCA-RISE-2018 JENNIFER2 grant agreement no.822070, H2020-MSCA-RISE-2019 SK2HK grant agreement no. 872549; 
and European Union's Next Generation EU/PRTR  grant CA3/RSUE2021-00559; 
the National Institute for Nuclear Physics (INFN), Italy.

The related data are released in~\cite{data-release}, where the detection time and the energy of decay electrons and neutrons are available for each selected stopping muons.

\bibliography{MuCapRefs}

\providecommand{\noopsort}[1]{}\providecommand{\singleletter}[1]{#1}%
\begin{thebibliography}{27}%
\makeatletter
\providecommand \@ifxundefined [1]{%
 \@ifx{#1\undefined}
}%
\providecommand \@ifnum [1]{%
 \ifnum #1\expandafter \@firstoftwo
 \else \expandafter \@secondoftwo
 \fi
}%
\providecommand \@ifx [1]{%
 \ifx #1\expandafter \@firstoftwo
 \else \expandafter \@secondoftwo
 \fi
}%
\providecommand \natexlab [1]{#1}%
\providecommand \enquote  [1]{``#1''}%
\providecommand \bibnamefont  [1]{#1}%
\providecommand \bibfnamefont [1]{#1}%
\providecommand \citenamefont [1]{#1}%
\providecommand \href@noop [0]{\@secondoftwo}%
\providecommand \href [0]{\begingroup \@sanitize@url \@href}%
\providecommand \@href[1]{\@@startlink{#1}\@@href}%
\providecommand \@@href[1]{\endgroup#1\@@endlink}%
\providecommand \@sanitize@url [0]{\catcode `\\12\catcode `\$12\catcode
  `\&12\catcode `\#12\catcode `\^12\catcode `\_12\catcode `\%12\relax}%
\providecommand \@@startlink[1]{}%
\providecommand \@@endlink[0]{}%
\providecommand \url  [0]{\begingroup\@sanitize@url \@url }%
\providecommand \@url [1]{\endgroup\@href {#1}{\urlprefix }}%
\providecommand \urlprefix  [0]{URL }%
\providecommand \Eprint [0]{\href }%
\providecommand \doibase [0]{https://doi.org/}%
\providecommand \selectlanguage [0]{\@gobble}%
\providecommand \bibinfo  [0]{\@secondoftwo}%
\providecommand \bibfield  [0]{\@secondoftwo}%
\providecommand \translation [1]{[#1]}%
\providecommand \BibitemOpen [0]{}%
\providecommand \bibitemStop [0]{}%
\providecommand \bibitemNoStop [0]{.\EOS\space}%
\providecommand \EOS [0]{\spacefactor3000\relax}%
\providecommand \BibitemShut  [1]{\csname bibitem#1\endcsname}%
\let\auto@bib@innerbib\@empty
\bibitem [{\citenamefont {Harada}\ \emph {et~al.}(2023)\citenamefont {Harada}
  \emph {et~al.}}]{Harada2023}%
  \BibitemOpen
  \bibfield  {author} {\bibinfo {author} {\bibfnamefont {M.}~\bibnamefont
  {Harada}} \emph {et~al.} (\bibinfo {collaboration} {Super-Kamiokande
  Collaboration}),\ }\bibfield  {title} {\bibinfo {title} {Search for
  astrophysical electron antineutrinos in {{Super-Kamiokande}} with 0.01\%
  gadolinium-loaded water},\ }\href {https://doi.org/10.3847/2041-8213/acdc9e}
  {\bibfield  {journal} {\bibinfo  {journal} {The Astrophysical Journal
  Letters}\ }\textbf {\bibinfo {volume} {951}},\ \bibinfo {pages} {L27}
  (\bibinfo {year} {2023})}\BibitemShut {NoStop}%
\bibitem [{\citenamefont {Takenaka}\ \emph {et~al.}(2020)\citenamefont
  {Takenaka} \emph {et~al.}}]{Takenaka2020a}%
  \BibitemOpen
  \bibfield  {author} {\bibinfo {author} {\bibfnamefont {A.}~\bibnamefont
  {Takenaka}} \emph {et~al.} (\bibinfo {collaboration} {Super-Kamiokande
  Collaboration}),\ }\bibfield  {title} {\bibinfo {title} {Search for proton
  decay via $p\rightarrow e^+ \pi^0$ and $p\rightarrow \mu^+ \pi^0$ with an
  enlarged fiducial volume in {Super}-{Kamiokande} {I}-{IV}},\ }\href
  {https://doi.org/10.1103/PhysRevD.102.112011} {\bibfield  {journal} {\bibinfo
   {journal} {Physical Review D}\ }\textbf {\bibinfo {volume} {102}},\ \bibinfo
  {pages} {112011} (\bibinfo {year} {2020})}\BibitemShut {NoStop}%
\bibitem [{\citenamefont {Wester}\ \emph {et~al.}(2024)\citenamefont {Wester}
  \emph {et~al.}}]{Wester2024}%
  \BibitemOpen
  \bibfield  {author} {\bibinfo {author} {\bibfnamefont {T.}~\bibnamefont
  {Wester}} \emph {et~al.} (\bibinfo {collaboration} {Super-Kamiokande
  Collaboration}),\ }\bibfield  {title} {\bibinfo {title} {Atmospheric neutrino
  oscillation analysis with neutron tagging and an expanded fiducial volume in
  {Super}-{Kamiokande} {I}--{V}},\ }\href
  {https://doi.org/10.1103/PhysRevD.109.072014} {\bibfield  {journal} {\bibinfo
   {journal} {Physical Review D}\ }\textbf {\bibinfo {volume} {109}},\ \bibinfo
  {pages} {072014} (\bibinfo {year} {2024})}\BibitemShut {NoStop}%
\bibitem [{\citenamefont {Suzuki}\ \emph {et~al.}(1987)\citenamefont {Suzuki},
  \citenamefont {Measday},\ and\ \citenamefont {Roalsvig}}]{Suzuki1987}%
  \BibitemOpen
  \bibfield  {author} {\bibinfo {author} {\bibfnamefont {T.}~\bibnamefont
  {Suzuki}}, \bibinfo {author} {\bibfnamefont {D.~F.}\ \bibnamefont
  {Measday}},\ and\ \bibinfo {author} {\bibfnamefont {J.~P.}\ \bibnamefont
  {Roalsvig}},\ }\bibfield  {title} {\bibinfo {title} {Total nuclear capture
  rates for negative muons},\ }\href {https://doi.org/10.1103/PhysRevC.35.2212}
  {\bibfield  {journal} {\bibinfo  {journal} {Physical Review C}\ }\textbf
  {\bibinfo {volume} {35}},\ \bibinfo {pages} {2212} (\bibinfo {year}
  {1987})}\BibitemShut {NoStop}%
\bibitem [{\citenamefont {Measday}(2001)}]{Measday2001}%
  \BibitemOpen
  \bibfield  {author} {\bibinfo {author} {\bibfnamefont {D.~F.}\ \bibnamefont
  {Measday}},\ }\bibfield  {title} {\bibinfo {title} {The nuclear physics of
  muon capture},\ }\href {https://doi.org/10.1016/S0370-1573(01)00012-6}
  {\bibfield  {journal} {\bibinfo  {journal} {Physics Report}\ }\textbf
  {\bibinfo {volume} {354}},\ \bibinfo {pages} {243} (\bibinfo {year}
  {2001})}\BibitemShut {NoStop}%
\bibitem [{\citenamefont {Kaplan}\ \emph {et~al.}(1958)\citenamefont {Kaplan},
  \citenamefont {Moyer},\ and\ \citenamefont {Pyle}}]{Kaplan1958}%
  \BibitemOpen
  \bibfield  {author} {\bibinfo {author} {\bibfnamefont {S.~N.}\ \bibnamefont
  {Kaplan}}, \bibinfo {author} {\bibfnamefont {B.~J.}\ \bibnamefont {Moyer}},\
  and\ \bibinfo {author} {\bibfnamefont {R.~V.}\ \bibnamefont {Pyle}},\
  }\bibfield  {title} {\bibinfo {title} {Neutron {{Emission Following}}
  $\ensuremath{\mu}$-{{Meson Capture}} in {{Silver}} and {{Lead}}},\ }\href
  {https://doi.org/10.1103/PhysRev.112.968} {\bibfield  {journal} {\bibinfo
  {journal} {Physical Review}\ }\textbf {\bibinfo {volume} {112}},\ \bibinfo
  {pages} {968} (\bibinfo {year} {1958})}\BibitemShut {NoStop}%
\bibitem [{\citenamefont {Winsberg}(1954)}]{Winsberg1954}%
  \BibitemOpen
  \bibfield  {author} {\bibinfo {author} {\bibfnamefont {L.}~\bibnamefont
  {Winsberg}},\ }\bibfield  {title} {\bibinfo {title} {Interaction of
  {{Negative Muons}} with {{Iodine}}},\ }\href
  {https://doi.org/10.1103/PhysRev.95.205} {\bibfield  {journal} {\bibinfo
  {journal} {Physical Review}\ }\textbf {\bibinfo {volume} {95}},\ \bibinfo
  {pages} {205} (\bibinfo {year} {1954})}\BibitemShut {NoStop}%
\bibitem [{\citenamefont {Hashim}\ \emph {et~al.}(2018)\citenamefont {Hashim},
  \citenamefont {Ejiri}, \citenamefont {Shima}, \citenamefont {Takahisa},
  \citenamefont {Sato}, \citenamefont {Kuno}, \citenamefont {Ninomiya},
  \citenamefont {Kawamura},\ and\ \citenamefont {Miyake}}]{Hashim2018}%
  \BibitemOpen
  \bibfield  {author} {\bibinfo {author} {\bibfnamefont {I.~H.}\ \bibnamefont
  {Hashim}}, \bibinfo {author} {\bibfnamefont {H.}~\bibnamefont {Ejiri}},
  \bibinfo {author} {\bibfnamefont {T.}~\bibnamefont {Shima}}, \bibinfo
  {author} {\bibfnamefont {K.}~\bibnamefont {Takahisa}}, \bibinfo {author}
  {\bibfnamefont {A.}~\bibnamefont {Sato}}, \bibinfo {author} {\bibfnamefont
  {Y.}~\bibnamefont {Kuno}}, \bibinfo {author} {\bibfnamefont {K.}~\bibnamefont
  {Ninomiya}}, \bibinfo {author} {\bibfnamefont {N.}~\bibnamefont {Kawamura}},\
  and\ \bibinfo {author} {\bibfnamefont {Y.}~\bibnamefont {Miyake}},\
  }\bibfield  {title} {\bibinfo {title} {Muon capture reaction on
  ${}^{100}${{Mo}} to study the nuclear response for double- $\beta$ decay and
  neutrinos of astrophysics origin},\ }\href
  {https://doi.org/10.1103/PhysRevC.97.014617} {\bibfield  {journal} {\bibinfo
  {journal} {Physical Review C}\ }\textbf {\bibinfo {volume} {97}},\ \bibinfo
  {pages} {1} (\bibinfo {year} {2018})}\BibitemShut {NoStop}%
\bibitem [{\citenamefont {Hashim}\ \emph {et~al.}(2020)\citenamefont {Hashim},
  \citenamefont {Ejiri}, \citenamefont {Othman}, \citenamefont {Ibrahim},
  \citenamefont {Soberi}, \citenamefont {Ghani}, \citenamefont {Shima},
  \citenamefont {Sato},\ and\ \citenamefont {Ninomiya}}]{Hashim2020}%
  \BibitemOpen
  \bibfield  {author} {\bibinfo {author} {\bibfnamefont {I.~H.}\ \bibnamefont
  {Hashim}}, \bibinfo {author} {\bibfnamefont {H.}~\bibnamefont {Ejiri}},
  \bibinfo {author} {\bibfnamefont {F.}~\bibnamefont {Othman}}, \bibinfo
  {author} {\bibfnamefont {F.}~\bibnamefont {Ibrahim}}, \bibinfo {author}
  {\bibfnamefont {F.}~\bibnamefont {Soberi}}, \bibinfo {author} {\bibfnamefont
  {N.~N.}\ \bibnamefont {Ghani}}, \bibinfo {author} {\bibfnamefont
  {T.}~\bibnamefont {Shima}}, \bibinfo {author} {\bibfnamefont
  {A.}~\bibnamefont {Sato}},\ and\ \bibinfo {author} {\bibfnamefont
  {K.}~\bibnamefont {Ninomiya}},\ }\bibfield  {title} {\bibinfo {title}
  {Nuclear isotope production by ordinary muon capture reaction},\ }\href
  {https://doi.org/10.1016/j.nima.2020.163749} {\bibfield  {journal} {\bibinfo
  {journal} {Nuclear Instruments and Methods in Physics Research A}\ }\textbf
  {\bibinfo {volume} {963}},\ \bibinfo {pages} {163749} (\bibinfo {year}
  {2020})}\BibitemShut {NoStop}%
\bibitem [{\citenamefont {Othman}\ \emph {et~al.}(2021)\citenamefont {Othman},
  \citenamefont {Hashim}, \citenamefont {Ejiri}, \citenamefont {Razali},
  \citenamefont {Ibrahim},\ and\ \citenamefont {Soberi}}]{Othman2021a}%
  \BibitemOpen
  \bibfield  {author} {\bibinfo {author} {\bibfnamefont {F.}~\bibnamefont
  {Othman}}, \bibinfo {author} {\bibfnamefont {I.}~\bibnamefont {Hashim}},
  \bibinfo {author} {\bibfnamefont {H.}~\bibnamefont {Ejiri}}, \bibinfo
  {author} {\bibfnamefont {R.}~\bibnamefont {Razali}}, \bibinfo {author}
  {\bibfnamefont {F.}~\bibnamefont {Ibrahim}},\ and\ \bibinfo {author}
  {\bibfnamefont {F.}~\bibnamefont {Soberi}},\ }\bibfield  {title} {\bibinfo
  {title} {Proton neutron emission model for muon charge exchange reaction},\
  }\bibfield  {journal} {\bibinfo  {journal} {American Institute of Physics
  Conference Series}\ }\href {https://doi.org/10.1063/5.0037769}
  {10.1063/5.0037769} (\bibinfo {year} {2021})\BibitemShut {NoStop}%
\bibitem [{\citenamefont {Kane}\ \emph {et~al.}(1973)\citenamefont {Kane},
  \citenamefont {Eckhause}, \citenamefont {Miller}, \citenamefont {Roberts},
  \citenamefont {Vislay},\ and\ \citenamefont {Welsh}}]{Kane1973}%
  \BibitemOpen
  \bibfield  {author} {\bibinfo {author} {\bibfnamefont {F.~R.}\ \bibnamefont
  {Kane}}, \bibinfo {author} {\bibfnamefont {M.}~\bibnamefont {Eckhause}},
  \bibinfo {author} {\bibfnamefont {G.~H.}\ \bibnamefont {Miller}}, \bibinfo
  {author} {\bibfnamefont {B.~L.}\ \bibnamefont {Roberts}}, \bibinfo {author}
  {\bibfnamefont {M.~E.}\ \bibnamefont {Vislay}},\ and\ \bibinfo {author}
  {\bibfnamefont {R.~E.}\ \bibnamefont {Welsh}},\ }\bibfield  {title} {\bibinfo
  {title} {Muon capture rates on ${}^{16}\mathrm{O}$ leading to bound states of
  ${}^{16}\mathrm{N}^*$},\ }\href
  {https://doi.org/10.1016/0370-2693(73)90206-2} {\bibfield  {journal}
  {\bibinfo  {journal} {Physics Letters B}\ }\textbf {\bibinfo {volume} {45}},\
  \bibinfo {pages} {292} (\bibinfo {year} {1973})}\BibitemShut {NoStop}%
\bibitem [{\citenamefont {Van Der~Schaaf}\ \emph {et~al.}(1983)\citenamefont
  {Van Der~Schaaf}, \citenamefont {Hermes}, \citenamefont {Powers},
  \citenamefont {Schlep{\"u}tz}, \citenamefont {Winter}, \citenamefont
  {Zglinski}, \citenamefont {Kozlowski}, \citenamefont {Bertl}, \citenamefont
  {Felawka}, \citenamefont {Hesselink},\ and\ \citenamefont {Van
  Der~Pluym}}]{VanDerSchaaf1983}%
  \BibitemOpen
  \bibfield  {author} {\bibinfo {author} {\bibfnamefont {A.}~\bibnamefont {Van
  Der~Schaaf}}, \bibinfo {author} {\bibfnamefont {E.~A.}\ \bibnamefont
  {Hermes}}, \bibinfo {author} {\bibfnamefont {R.~J.}\ \bibnamefont {Powers}},
  \bibinfo {author} {\bibfnamefont {F.~W.}\ \bibnamefont {Schlep{\"u}tz}},
  \bibinfo {author} {\bibfnamefont {R.~G.}\ \bibnamefont {Winter}}, \bibinfo
  {author} {\bibfnamefont {A.}~\bibnamefont {Zglinski}}, \bibinfo {author}
  {\bibfnamefont {T.}~\bibnamefont {Kozlowski}}, \bibinfo {author}
  {\bibfnamefont {W.}~\bibnamefont {Bertl}}, \bibinfo {author} {\bibfnamefont
  {L.}~\bibnamefont {Felawka}}, \bibinfo {author} {\bibfnamefont {W.~H.}\
  \bibnamefont {Hesselink}},\ and\ \bibinfo {author} {\bibfnamefont
  {J.}~\bibnamefont {Van Der~Pluym}},\ }\bibfield  {title} {\bibinfo {title}
  {Measurement of neutron energy spectra and neutron-gamma angular correlations
  for the muon capture process ${}^{16}\mathrm{O}(\mu^-, \nu_\mu xn){}^{14,
  15}\mathrm{N}$},\ }\href {https://doi.org/10.1016/0375-9474(83)90246-4}
  {\bibfield  {journal} {\bibinfo  {journal} {Nuclear Physics A}\ }\textbf
  {\bibinfo {volume} {408}},\ \bibinfo {pages} {573} (\bibinfo {year}
  {1983})}\BibitemShut {NoStop}%
\bibitem [{\citenamefont {Maekawa}\ \emph {et~al.}(2025)\citenamefont {Maekawa}
  \emph {et~al.}}]{maekawa2025}%
  \BibitemOpen
  \bibfield  {author} {\bibinfo {author} {\bibfnamefont {Y.}~\bibnamefont
  {Maekawa}} \emph {et~al.},\ }\bibfield  {title} {\bibinfo {title}
  {Measurement of the branching ratio of $\mathrm{^{16}N}$, $\mathrm{^{15}C}$,
  $\mathrm{^{12}B}$, and $\mathrm{^{13}B}$ isotopes through the nuclear muon
  capture reaction in the super-kamiokande detector},\ }\href
  {https://doi.org/10.1103/b93l-dbpj} {\bibfield  {journal} {\bibinfo
  {journal} {Physical Review C}\ }\textbf {\bibinfo {volume} {112}},\ \bibinfo
  {pages} {064614} (\bibinfo {year} {2025})}\BibitemShut {NoStop}%
\bibitem [{\citenamefont {Abe}\ \emph {et~al.}(2022{\natexlab{a}})\citenamefont
  {Abe} \emph {et~al.}}]{Abe2022}%
  \BibitemOpen
  \bibfield  {author} {\bibinfo {author} {\bibfnamefont {K.}~\bibnamefont
  {Abe}} \emph {et~al.} (\bibinfo {collaboration} {Super-Kamiokande
  Collaboration}),\ }\bibfield  {title} {\bibinfo {title} {First {{Gadolinium
  Loading}} to {{Super-Kamiokande}}},\ }\href
  {https://doi.org/10.1016/j.nima.2021.166248} {\bibfield  {journal} {\bibinfo
  {journal} {Nuclear Inst. and Methods in Physics Research, A}\ }\textbf
  {\bibinfo {volume} {1027}},\ \bibinfo {pages} {166248} (\bibinfo {year}
  {2022}{\natexlab{a}})}\BibitemShut {NoStop}%
\bibitem [{\citenamefont {Fukuda}\ \emph {et~al.}(2003)\citenamefont {Fukuda}
  \emph {et~al.}}]{Fukuda2003}%
  \BibitemOpen
  \bibfield  {author} {\bibinfo {author} {\bibfnamefont {S.}~\bibnamefont
  {Fukuda}} \emph {et~al.} (\bibinfo {collaboration} {Super-Kamiokande
  Collaboration}),\ }\bibfield  {title} {\bibinfo {title} {The
  {{Super-Kamiokande}} detector},\ }\href
  {https://doi.org/10.1016/S0168-9002(03)00425-X} {\bibfield  {journal}
  {\bibinfo  {journal} {Nuclear Instruments and Methods in Physics Research A}\
  }\textbf {\bibinfo {volume} {501}},\ \bibinfo {pages} {418} (\bibinfo {year}
  {2003})}\BibitemShut {NoStop}%
\bibitem [{\citenamefont {Abe}\ \emph {et~al.}(2014)\citenamefont {Abe} \emph
  {et~al.}}]{Abe2014}%
  \BibitemOpen
  \bibfield  {author} {\bibinfo {author} {\bibfnamefont {K.}~\bibnamefont
  {Abe}} \emph {et~al.},\ }\bibfield  {title} {\bibinfo {title} {Calibration of
  the {{Super-Kamiokande}} detector},\ }\href
  {https://doi.org/10.1016/j.nima.2013.11.081} {\bibfield  {journal} {\bibinfo
  {journal} {Nuclear Instruments and Methods in Physics Research A}\ }\textbf
  {\bibinfo {volume} {737}},\ \bibinfo {pages} {253} (\bibinfo {year}
  {2014})}\BibitemShut {NoStop}%
\bibitem [{\citenamefont {Zyla}\ \emph {et~al.}(2020)\citenamefont {Zyla} \emph
  {et~al.}}]{PDG2020}%
  \BibitemOpen
  \bibfield  {author} {\bibinfo {author} {\bibfnamefont {P.}~\bibnamefont
  {Zyla}} \emph {et~al.} (\bibinfo {collaboration} {Particle Data Group}),\
  }\bibfield  {title} {\bibinfo {title} {Review of {{Particle Physics}}},\
  }\href {https://doi.org/10.1093/ptep/ptaa104} {\bibfield  {journal} {\bibinfo
   {journal} {PTEP}\ }\textbf {\bibinfo {volume} {2020}},\ \bibinfo {pages}
  {083C01} (\bibinfo {year} {2020})}\BibitemShut {NoStop}%
\bibitem [{\citenamefont {Nishino}\ \emph {et~al.}(2009)\citenamefont
  {Nishino}, \citenamefont {Awai}, \citenamefont {Hayato}, \citenamefont
  {Nakayama}, \citenamefont {Okumura}, \citenamefont {Shiozawa}, \citenamefont
  {Takeda}, \citenamefont {Ishikawa}, \citenamefont {Minegishi},\ and\
  \citenamefont {Arai}}]{Nishino2009}%
  \BibitemOpen
  \bibfield  {author} {\bibinfo {author} {\bibfnamefont {H.}~\bibnamefont
  {Nishino}}, \bibinfo {author} {\bibfnamefont {K.}~\bibnamefont {Awai}},
  \bibinfo {author} {\bibfnamefont {Y.}~\bibnamefont {Hayato}}, \bibinfo
  {author} {\bibfnamefont {S.}~\bibnamefont {Nakayama}}, \bibinfo {author}
  {\bibfnamefont {K.}~\bibnamefont {Okumura}}, \bibinfo {author} {\bibfnamefont
  {M.}~\bibnamefont {Shiozawa}}, \bibinfo {author} {\bibfnamefont
  {A.}~\bibnamefont {Takeda}}, \bibinfo {author} {\bibfnamefont
  {K.}~\bibnamefont {Ishikawa}}, \bibinfo {author} {\bibfnamefont
  {A.}~\bibnamefont {Minegishi}},\ and\ \bibinfo {author} {\bibfnamefont
  {Y.}~\bibnamefont {Arai}},\ }\bibfield  {title} {\bibinfo {title} {High-speed
  charge-to-time converter {{ASIC}} for the {{Super-Kamiokande}} detector},\
  }\href {https://doi.org/10.1016/j.nima.2009.09.026} {\bibfield  {journal}
  {\bibinfo  {journal} {Nuclear Instruments and Methods in Physics Research A}\
  }\textbf {\bibinfo {volume} {610}},\ \bibinfo {pages} {710} (\bibinfo {year}
  {2009})}\BibitemShut {NoStop}%
\bibitem [{\citenamefont {Nakano}\ \emph {et~al.}(2020)\citenamefont {Nakano},
  \citenamefont {Hokama}, \citenamefont {Matsubara}, \citenamefont {Miwa},
  \citenamefont {Nakahata}, \citenamefont {Nakamura}, \citenamefont {Sekiya},
  \citenamefont {Takeuchi}, \citenamefont {Tasaka},\ and\ \citenamefont
  {Wendell}}]{Nakano2020}%
  \BibitemOpen
  \bibfield  {author} {\bibinfo {author} {\bibfnamefont {Y.}~\bibnamefont
  {Nakano}}, \bibinfo {author} {\bibfnamefont {T.}~\bibnamefont {Hokama}},
  \bibinfo {author} {\bibfnamefont {M.}~\bibnamefont {Matsubara}}, \bibinfo
  {author} {\bibfnamefont {M.}~\bibnamefont {Miwa}}, \bibinfo {author}
  {\bibfnamefont {M.}~\bibnamefont {Nakahata}}, \bibinfo {author}
  {\bibfnamefont {T.}~\bibnamefont {Nakamura}}, \bibinfo {author}
  {\bibfnamefont {H.}~\bibnamefont {Sekiya}}, \bibinfo {author} {\bibfnamefont
  {Y.}~\bibnamefont {Takeuchi}}, \bibinfo {author} {\bibfnamefont
  {S.}~\bibnamefont {Tasaka}},\ and\ \bibinfo {author} {\bibfnamefont {R.~A.}\
  \bibnamefont {Wendell}},\ }\bibfield  {title} {\bibinfo {title} {Measurement
  of the radon concentration in purified water in the {{Super-Kamiokande IV}}
  detector},\ }\href {https://doi.org/10.1016/j.nima.2020.164297} {\bibfield
  {journal} {\bibinfo  {journal} {Nuclear Instruments and Methods in Physics
  Research A}\ }\textbf {\bibinfo {volume} {977}},\ \bibinfo {pages} {164297}
  (\bibinfo {year} {2020})}\BibitemShut {NoStop}%
\bibitem [{\citenamefont {Zhang}\ \emph {et~al.}(2016)\citenamefont {Zhang}
  \emph {et~al.}}]{Zhang2016}%
  \BibitemOpen
  \bibfield  {author} {\bibinfo {author} {\bibfnamefont {Y.}~\bibnamefont
  {Zhang}} \emph {et~al.} (\bibinfo {collaboration} {Super-Kamiokande
  Collaboration}),\ }\bibfield  {title} {\bibinfo {title} {First measurement of
  radioactive isotope production through cosmic-ray muon spallation in
  {{Super-Kamiokande IV}}},\ }\href
  {https://doi.org/10.1103/PhysRevD.93.012004} {\bibfield  {journal} {\bibinfo
  {journal} {Physical Review D}\ }\textbf {\bibinfo {volume} {93}},\ \bibinfo
  {pages} {012004} (\bibinfo {year} {2016})}\BibitemShut {NoStop}%
\bibitem [{\citenamefont {Kitagawa}\ \emph {et~al.}(2024)\citenamefont
  {Kitagawa} \emph {et~al.}}]{Kitagawa2024}%
  \BibitemOpen
  \bibfield  {author} {\bibinfo {author} {\bibfnamefont {H.}~\bibnamefont
  {Kitagawa}} \emph {et~al.} (\bibinfo {collaboration} {Super-Kamiokande
  Collaboration}),\ }\bibfield  {title} {\bibinfo {title} {Measurements of the
  charge ratio and polarization of cosmic-ray muons with the
  {{Super-Kamiokande}} detector},\ }\href
  {https://doi.org/10.1103/PhysRevD.110.082008} {\bibfield  {journal} {\bibinfo
   {journal} {Physical Review D}\ }\textbf {\bibinfo {volume} {110}},\ \bibinfo
  {pages} {082008} (\bibinfo {year} {2024})}\BibitemShut {NoStop}%
\bibitem [{\citenamefont {Han}\ \emph {et~al.}(2025)\citenamefont {Han} \emph
  {et~al.}}]{Han2025}%
  \BibitemOpen
  \bibfield  {author} {\bibinfo {author} {\bibfnamefont {S.}~\bibnamefont
  {Han}} \emph {et~al.} (\bibinfo {collaboration} {Super-Kamiokande
  Collaboration}),\ }\bibfield  {title} {\bibinfo {title} {Measurement of
  neutron production in atmospheric neutrino interactions at
  {{Super-Kamiokande}}},\ }\href {https://doi.org/10.1103/4d71-d69k} {\bibfield
   {journal} {\bibinfo  {journal} {Physical Review D}\ }\textbf {\bibinfo
  {volume} {112}},\ \bibinfo {pages} {012004} (\bibinfo {year}
  {2025})}\BibitemShut {NoStop}%
\bibitem [{\citenamefont {Han}(2023)}]{Han2023}%
  \BibitemOpen
  \bibfield  {author} {\bibinfo {author} {\bibfnamefont {S.}~\bibnamefont
  {Han}},\ }\emph {\bibinfo {title} {Measurement of Neutrons Produced in
  Atmospheric Neutrino Interactions at {{Super-Kamiokande}}}},\ \href@noop {}
  {Ph.D. thesis},\ \bibinfo  {school} {University of Tokyo} (\bibinfo {year}
  {2023})\BibitemShut {NoStop}%
\bibitem [{\citenamefont {Abe}\ \emph {et~al.}(2022{\natexlab{b}})\citenamefont
  {Abe} \emph {et~al.}}]{Abe2022a}%
  \BibitemOpen
  \bibfield  {author} {\bibinfo {author} {\bibfnamefont {K.}~\bibnamefont
  {Abe}} \emph {et~al.} (\bibinfo {collaboration} {Super-Kamiokande
  Collaboration}),\ }\bibfield  {title} {\bibinfo {title} {Neutron tagging
  following atmospheric neutrino events in a water {{Cherenkov}} detector},\
  }\href {https://doi.org/10.1088/1748-0221/17/10/P10029} {\bibfield  {journal}
  {\bibinfo  {journal} {Journal of Instrumentation}\ }\textbf {\bibinfo
  {volume} {17}}\bibinfo  {number} { (10)},\ \bibinfo {pages}
  {P10029}}\BibitemShut {NoStop}%
\bibitem [{\citenamefont {Feldman}\ and\ \citenamefont
  {Cousins}(1998)}]{Feldman1998}%
  \BibitemOpen
\bibfield  {number} {  }\bibfield  {author} {\bibinfo {author} {\bibfnamefont
  {G.~J.}\ \bibnamefont {Feldman}}\ and\ \bibinfo {author} {\bibfnamefont
  {R.~D.}\ \bibnamefont {Cousins}},\ }\bibfield  {title} {\bibinfo {title}
  {Unified approach to the classical statistical analysis of small signals},\
  }\href {https://doi.org/10.1103/PhysRevD.57.3873} {\bibfield  {journal}
  {\bibinfo  {journal} {Physical Review D}\ }\textbf {\bibinfo {volume} {57}},\
  \bibinfo {pages} {3873} (\bibinfo {year} {1998})}\BibitemShut {NoStop}%
\bibitem [{\citenamefont {Von~Egidy}\ and\ \citenamefont
  {Hartmann}(1982)}]{VonEgidy1982}%
  \BibitemOpen
  \bibfield  {author} {\bibinfo {author} {\bibfnamefont {T.}~\bibnamefont
  {Von~Egidy}}\ and\ \bibinfo {author} {\bibfnamefont {F.~J.}\ \bibnamefont
  {Hartmann}},\ }\bibfield  {title} {\bibinfo {title} {Average muonic
  {{Coulomb}} capture probabilities for 65 elements},\ }\href
  {https://doi.org/10.1103/PhysRevA.26.2355} {\bibfield  {journal} {\bibinfo
  {journal} {Physical Review A}\ }\textbf {\bibinfo {volume} {26}},\ \bibinfo
  {pages} {2355} (\bibinfo {year} {1982})}\BibitemShut {NoStop}%
\bibitem [{dat(2025)}]{data-release}%
  \BibitemOpen
  \bibfield  {title} {\bibinfo {title} {Data release: Neutron multiplicity
  measurement in muon capture on oxygen nuclei in the {Gd-loaded}
  {Super-Kamiokande} detector},\ }\href
  {https://doi.org/10.5281/zenodo.15081911} {10.5281/zenodo.15081911} (\bibinfo
  {year} {2025})\BibitemShut {NoStop}%
\end{thebibliography}%

\end{document}